\documentclass[preprint,11pt]{elsarticle}
\usepackage[a4paper,margin=1in]{geometry}

\usepackage[T1]{fontenc} 
\usepackage[normalem]{ulem} 

\usepackage[dvipsnames,svgnames,x11names]{xcolor}
\usepackage{listings}

\lstdefinelanguage{XML}
{
basicstyle=\ttfamily\footnotesize,
  morestring=[b]",
  moredelim=[s][\bfseries\color{Maroon}]{<}{\ },
  moredelim=[s][\bfseries\color{Maroon}]{</}{>},
  moredelim=[l][\bfseries\color{Maroon}]{/>},
  moredelim=[l][\bfseries\color{Maroon}]{>},
  morecomment=[s]{<?}{?>},
  morecomment=[s]{<!--}{-->},
  commentstyle=\color{gray},
  stringstyle=\color{blue},
  identifierstyle=\color{red}
}
\usepackage{moreverb}

\usepackage[nounderscore]{syntax}

\usepackage[cmex10]{amsmath}
\usepackage{amssymb}
\usepackage{mathtools}
\usepackage{amsthm}
\usepackage{amsfonts}

\usepackage{gensymb}

\usepackage{subfig} 
\usepackage[export]{adjustbox}

\usepackage{algorithmicx}
\usepackage{algpseudocode}
\usepackage[ruled]{algorithm}
\definecolor{light-gray}{gray}{0.75}
\definecolor{dark-gray}{gray}{0.5}
\algrenewcommand{\algorithmiccomment}[1]{\hskip3em{{\footnotesize \textcolor{light-gray}{$\blacktriangleright$}~\textcolor{dark-gray}{#1}}}}
\usepackage{multirow} 
\usepackage{rotating} 
\usepackage{booktabs} 
\usepackage{colortbl} 
\usepackage{tablefootnote} 

\usepackage[pdftex,colorlinks=true,urlcolor=blue,citecolor=blue]{hyperref}

\usepackage{xspace}

\usepackage{enumitem}

\usepackage{balance}

\usepackage{blindtext}

\begin{document}

\begin{frontmatter}
\title{AeroResQ: Edge-Accelerated UAV Framework for Scalable, Resilient and Collaborative Escape Route Planning in Wildfire Scenarios}

\author[inst1]{Suman Raj\fnref{*}}
\fntext[*]{Correspondence to: USC Information Sciences Institute, 4676 Admiralty Way Suite 1001, Marina del Rey, CA, 90292, USA.}
\ead{sumanraj@iisc.ac.in}
\author[inst1]{Radhika Mittal}
\ead{ced19i050@iiitdm.ac.in}
\author[inst2]{Rajiv Mayani}
\ead{mayani@isi.edu}
\author[inst2]{Pawel Zuk}
\ead{pawel.m.zuk@gmail.com}
\author[inst3]{Anirban Mandal} 
\ead{anirban@renci.org}
\author[inst4]{Michael Zink}
\ead{zink@ecs.umass.edu}
\author[inst1]{Yogesh Simmhan}
\ead{simmhan@iisc.ac.in}
\author[inst2]{Ewa Deelman}
\ead{deelman@isi.edu}

\affiliation[inst1]{organization={Indian Institute of Science},
            addressline={CV Raman Road}, 
            city={Bengaluru},
            state={KA},
            country={India}}
\affiliation[inst2]{organization={Information Sciences Institute, University of Southern California},
            state={CA},
            country={USA}}
\affiliation[inst3]{organization={Renaissance Computing Institute, University of North Carolina },
            addressline={Chapel Hill}, 
            state={NC},
            country={USA}}
\affiliation[inst4]{organization={University of Massachusetts},
            city={Amherst},
            state={MA},
            country={USA}}

\begin{abstract}
Drone fleets equipped with onboard cameras, computer vision, and Deep Neural Network (DNN) models present a powerful paradigm for real-time spatio-temporal decision-making. In wildfire response, such drones play a pivotal role in monitoring fire dynamics, supporting firefighter coordination, and facilitating safe evacuation. In this paper, we introduce \textit{AeroResQ}, an edge-accelerated UAV framework designed for \textit{scalable, resilient, and collaborative escape route planning} during wildfire scenarios. AeroResQ adopts a \textit{multi-layer orchestration architecture} comprising \textit{service drones (SDs)} and \textit{coordinator drones (CDs)}, each performing specialized roles. SDs survey fire-affected areas, detect stranded individuals using onboard edge accelerators running fire detection and human pose identification DNN models, and issue requests for assistance.  CDs, equipped with lightweight data stores such as Apache IoTDB, dynamically generate optimal ground escape routes and monitor firefighter movements along these routes. The framework proposes a \textit{collaborative path-planning approach} based on a weighted A* search algorithm, where CDs compute \textit{context-aware escape paths} that adapt to evolving wildfire conditions and firefighter positions. AeroResQ further incorporates \textit{intelligent load-balancing} and \textit{resilience mechanisms}: CD failures trigger \textit{automated data redistribution} across IoTDB replicas, while SD failures initiate \textit{geo-fenced re-partitioning} and reassignment of spatial workloads to operational SDs. We evaluate AeroResQ using realistic wildfire emulated setup modeled on recent Southern California wildfires. Experimental results demonstrate that AeroResQ achieves a nominal end-to-end latency of $\leq 500$~ms, much below the 2s request interval, while maintaining over 98\% successful task reassignment and completion, underscoring its feasibility for \textit{real-time, on-field deployment} in emergency response and firefighter safety operations.
\end{abstract}


\end{frontmatter}

\section{Introduction} 

\paragraph{Context}
Unmanned Aerial Vehicles (UAVs), commonly known as drones, have gained significant prominence in recent years due to advancements in autonomous navigation, edge computing, and on-device Artificial Intelligence (AI). These developments have enabled UAVs to process data in real-time without relying on remote cloud infrastructure, making them highly efficient for mission-critical applications such as disaster management, environmental monitoring, and smart city operations. By leveraging AI-driven decision-making at the edge, UAVs can autonomously navigate complex environments, detect anomalies, and adapt to dynamic conditions with minimal human intervention.

\paragraph{Motivation} One of the most prominent applications of UAVs is in disaster response, where their ability to provide real-time situational awareness and assistance in decision-making has proven invaluable. Among various disaster scenarios, wildfires present a particularly challenging environment, demanding rapid detection, continuous monitoring, and coordinated evacuation efforts. \textit{Wildfires} spread unpredictably, often outpacing conventional firefighting strategies. Traditional aerial surveillance methods, such as manned helicopters and satellites, face limitations in terms of cost, operational flexibility, and response time. UAVs equipped with onboard AI models offer a transformative approach to wildfire response by enabling real-time fire detection, perimeter tracking, and intelligent escape route planning. Through the integration of Computer Vision (CV), Deep Neural Networks (DNNs), and edge accelerators, drones can efficiently identify fire hotspots, locate stranded individuals, and assist firefighters in navigating through hazardous terrain. This motivates us to use a fleet of drones that can enable scalable and coordinated operations, covering the entire swath of the wildfire area, ensuring comprehensive fire monitoring, efficient rescue missions, and real-time adaptive decision-making, even in rapidly evolving wildfire conditions.

\paragraph{Operational Considerations} Although drones are typically not allowed to fly during fire fighting operations, this ban primarily targets civilian drones, often flown by individuals who may have ulterior motives. In our work, we envision drones that will be deployed in coordination with other aerial and ground assets to make sure that they do not interfere with other operations.

\paragraph{Challenges} Wildfires pose significant challenges to emergency response teams, often requiring rapid and coordinated interventions to assist stranded individuals and firefighters (evacuees), and guide them toward safe locations. 
In disaster scenarios such as wildfires, relying on cloud communication poses significant challenges due to unreliable network connectivity, high-latency data transmission, and bandwidth constraints. Infrastructure damage or network congestion in affected areas can severely limit access to cloud resources, delaying critical decision-making processes for real-time situational awareness. 
\textit{Edge computing} leveraging onboard AI models mitigates these challenges by enabling UAVs to process and analyze data locally, reducing dependency on external connectivity and ensuring faster response times. Moreover, edge-based processing enhances system resilience, allowing UAVs to continue operating even if cloud services become inaccessible.

When deploying a fleet of drones for wildfire response, several critical challenges arise, particularly when drones fail due to battery depletion or exposure to wildfire heat. One of the primary challenges is \textit{coverage gaps in surveillance} when surveillance drones fail. If they fail, gaps in real-time situational awareness can occur, potentially leading to missed detections of individuals in distress. To mitigate this, the system must dynamically redistribute remaining drones or deploy replacements to ensure continuous monitoring of the affected region. Another significant concern because of drone failures is the \textit{loss of critical data} collected by these drones, including fire perimeter updates and human detection information. If a drone fails before offloading its data, escape route planning and evacuee safety could be compromised. Next, the impact on \textit{escape route generation} is another major issue. Losing one of the drones responsible for planning safe evacuation paths for evacuees, could delay route computation, increasing risk for those on the ground. The system must be designed with redundant computing capabilities, where another drone or some base station can seamlessly take over escape route generation without delays. \textit{Keeping track of drones responsible for monitoring evacuees} along escape routes to ensure that they reach safe zones is also critical. Their failure could result in a loss of tracking and guidance, making evacuees vulnerable to changing fire conditions. To prevent this, the system must support real-time handoff mechanisms, where another drone can immediately assume monitoring responsibilities if a drone is lost. 

Addressing these challenges requires a resilient wildfire drone swarm that integrates adaptive mission planning, fault-tolerant data replication, and intelligent task redistribution to ensure continuous operation, evacuee safety, and effective wildfire containment.

\paragraph{Gaps} Existing studies on UAV-based wildfire management primarily focus on fire detection, monitoring, and emergency response, with some leveraging UAVs for data collection and risk estimation. There is only limited research addressing the critical application of UAV-based escape route planning, with most approaches relying only on fire-spread models and lacking coordinated drone fleet deployment or resilience considerations. While prior works explore resilient multi-UAV coordination for continuous fire tracking and fault tolerance, they do not incorporate dynamic escape route planning for evacuees. 

\paragraph{Contributions} 
In this article, we present \textit{AeroResQ}, a novel drone fleet architecture that offers resilient and collaborative escape route planning for stranded individuals using edge-accelerated UAVs during wildfire scenarios. Specifically, we propose the following contributions: 

\begin{enumerate}[leftmargin=*]
    \item We motivate our research and outline the application and system requirements for AeroResQ, and introduce key components of AeroResQ, such as the Base Station (BS), Service Drones (SD), and Coordinator Drones (CD)(§~\ref{sec:problem-overview}).
    \item We propose AeroResQ, a novel execution workflow designed for real-time wildfire management and rescue assistance (§~\ref{sec:workflow}) that uses a weighted A* algorithm to generate the safest and shortest escape route (\S~\ref{sec:collaborative-planning}).
    \item We develop resilience algorithms that ensure system continuity in the event of service or coordinator drone failures, implementing fault-tolerant mechanisms to guarantee safe evacuation (§~\ref{sec:resilience-algorithms}).
    \item We describe the architecture and implementation details of AeroResQ, including the services deployed on drones, their communication framework, and the DNN models used for fire detection and human localization (§~\ref{sec:arch}).
    \item We evaluate AeroResQ using real wildfire datasets from recent California wildfires (2025), analyzing system performance for escape route planning and coordination under different emulated failure and dynamic scenarios, and demonstrating its effectiveness in large-scale deployments (§~\ref{sec:results}).
\end{enumerate}

We review related work (\S~\ref{sec:related}), discuss our findings (\S~\ref{sec:discussion}) and offer our conclusions and future work(\S~\ref{sec:conclusion}).

\section{Problem Overview}\label{sec:problem-overview}

\begin{figure}[t]
    \centering
    \includegraphics[width=0.85\columnwidth]{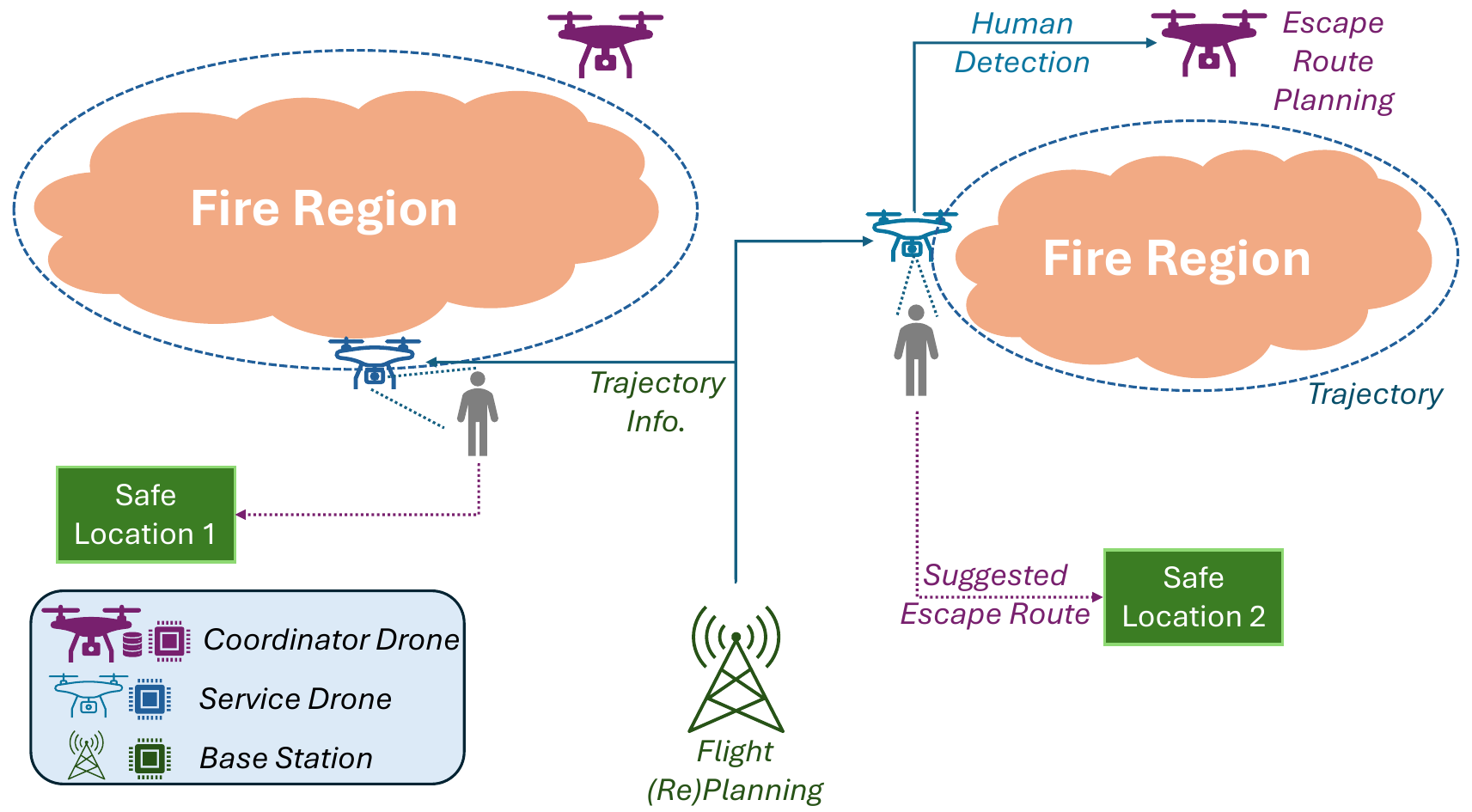}
    \caption{Problem Overview.}
    \label{fig:app}
\end{figure}

We provide an overview of the problem using Fig.~\ref{fig:app}, which showcases how different drone types collaborate for wildfire response and evacuation planning. The fire regions represent active wildfire zones where drones are deployed for monitoring and rescue operations. Service drones (blue) conduct aerial surveillance, detecting human presence (Fig.~\ref{fig:service-drone-view}) and assessing fire spread. They transmit trajectory information to the base station (green), which acts as the central command, overseeing flight planning and real-time updates. Coordinator drones (purple) play a crucial role in escape route planning by computing the safest evacuation paths based on fire coverage and terrain accessibility. Once a route is determined, the system guides evacuees toward safe locations.

\subsection{Application Requirements} 
We envision AeroResQ as a comprehensive drone-based system designed to support multiple critical applications in wildfire response by leveraging AI-driven decision-making and autonomous UAV operations. 
\textit{Wildfire monitoring} enables continuous aerial surveillance using onboard sensors and AI models to detect and track fire spread in real time, providing vital data for incident commanders. 
\textit{Escape route planning} should generate evacuation paths for both civilians and firefighters by incorporating fire spread information, terrain accessibility constraints, and potential safe zones. \textit{Evacuee tracking} plays a crucial role in identifying and localizing stranded individuals, ensuring that they receive timely assistance while navigating towards safe zones. This capability is essential for coordinating rescue efforts and minimizing risk in rapidly evolving wildfire scenarios. Additionally, \textit{resource allocation and coordination} optimize drone deployments and task assignments, ensuring that UAVs are utilized efficiently for surveillance, path planning, and rescue operations.

\subsection{System Requirements}
Such applications requires a robust combination of hardware, software, networking, and intelligent control mechanisms to effectively assist stranded individuals in wildfire management. Below, we outline the key system requirements.

\subsubsection{Hardware Requirements}
\begin{figure}
    \centering
    \subfloat[Service drone view]{
    \includegraphics[width=0.27\columnwidth]{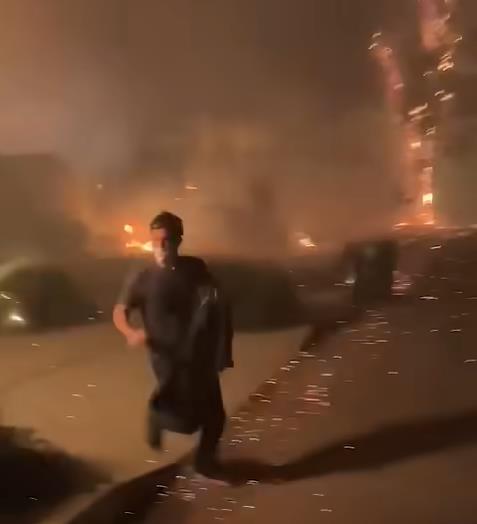}
    \label{fig:service-drone-view}
    }
    \subfloat[Coordinator drone view]{
    \includegraphics[width=0.3\columnwidth]{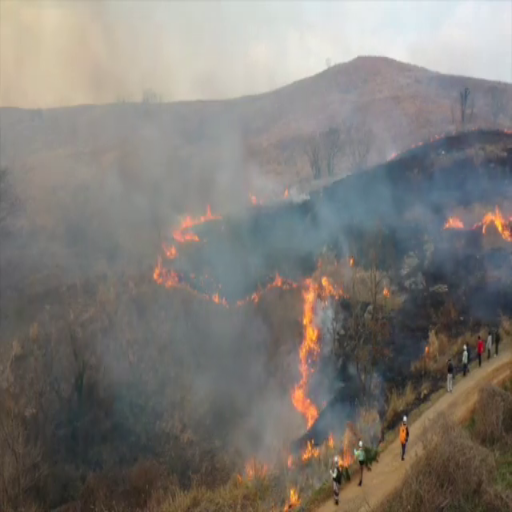}
    \label{fig:coordinator-drone-view}
    }  
    \caption{Sample images from drones: (a) Palisades fire (b) FlameVision\cite{Jafar2023FlameVision} 
    }
    \label{fig:drone-images}
\end{figure}

A fleet of UAVs (drones) equipped with specialized hardware to support real-time inferencing and communication is a critical requirement. The service drones must be lightweight yet capable of edge computing, that can process onboard deep learning models in real-time. The drones also require high-resolution optical and infrared cameras to capture geotagged imagery of fire and humans. Since the coordinator drones operate at a higher altitude to maintain a broader situational awareness (Fig.~\ref{fig:coordinator-drone-view}), they need powerful edge computing capabilities, in addition to the ones service drones provide. Since these help with escape route generation, they need to have high-capacity storage to store ground maps. 
The base stations must be equipped with high-performance computing servers, possibly with GPU accelerators, to support dynamic mission updates. Additionally, reliability is crucial, as the base station serves as the central coordination hub, ensuring seamless operation even in network disruptions or hardware failures.

\subsubsection{Software and AI-driven Intelligence} 
A system designed for wildfire management and firefighter assistance must integrate several critical components to ensure real-time perception, decision-making, and coordination. It requires lightweight and optimized deep learning models for fire detection, allowing dynamic updates to the fire perimeter, and human pose estimation, enabling the identification of stranded individuals. An adaptive flight planning module at the base station should be capable of dynamically assigning drones in case of service drone failure because of fire or battery depletion, while also modifying drone trajectories in real time to avoid hazardous zones. Safe evacuation path generation relies on shortest path algorithms such as weighted A* to determine efficient escape routes. Additionally, a lightweight distributed datastore, such as IoTDB, is essential for storing real-time evacuees location data, ensuring fast request retrieval and resilient data replication across coordinator drones to prevent data loss in case of drone failures. 

\subsubsection{Resilience and Fault Tolerance}
A wildfire response system must be resilient and fault-tolerant, ensuring continuous operation even in the face of failures. 
In the event of a service drone failure, the system should dynamically reallocate the fire-affected area to maintain surveillance and assistance coverage using the existing set of drones. Given the critical role of coordinator drones in route planning and data storage, their failure must be mitigated through data replication strategies, ensuring other drones retain active requests, and by dynamically assigning requests to remaining coordinator drones. Additionally, the system must prioritize energy-efficient operations through smart battery management to optimize drone activity and implement mid-mission handoff mechanisms, allowing one drone to seamlessly take over before another runs out of power.

\subsubsection{Networking and Communication}
A wildfire response system must ensure robust networking and communication to support real-time, low-latency coordination between drones and the base station. It should integrate high-bandwidth wireless protocols such as 5G, LTE, or dedicated mesh networks to facilitate seamless data exchange. Edge-to-cloud synchronization could also be employed to replicate critical data across drones and the base station, ensuring resilience. To enhance reliability in remote areas, the system should also incorporate redundant communication channels, such as satellite links, where cellular connectivity is limited. \\

\noindent While we focus on addressing the first three requirements in this paper, we do not dive deep into the networking part of the problem. This is outside the scope of this article and will be considered for future work.

\subsection{AeroResQ Components}
Now, we discuss the various components of AeroResQ that aims to address the above requirements.
\subsubsection{Drone Fleet Service Provider}
The entire fleet of drones is managed by a Drone Fleet Service Provider (DFSP), such as Skydio DFR. These providers are a part of wildfire management team such as CalFire, and work in conjunction with them to assist firefighters. The DFSP hosts a base station (BS), which is a reliable entity, like a mobile command center or an Unmanned Ground Vehicle.
The provider ensures the availability of drones with required computing resources, such as onboard edge 
accelerated processing units, and integrates required machine learning models. The DFSP also manages communication infrastructure to maintain seamless coordination among drones, ground firefighters, and command centers, ensuring effective mission execution.
They continuously monitor drone operations, overseeing battery recharging, replacements, and network resilience. In case of a drone failure, the DFSP dynamically reassigns tasks to maintain uninterrupted service and replans flight paths to adapt to changing wildfire conditions.

\subsubsection{Service Drones for Ground-Level Surveillance}
Service drones (SDs) are equipped with onboard edge-
accelerated computing devices to facilitate real-time computer vision tasks. The onboard cameras on these drones capture images and video feeds, which are processed onboard using DNN models for various tasks. Specifically, a drone hosting a \textit{human detection} DNN~\cite{NVIDIA_TRTPose} can identify individuals, and if they require assistance, relay their geospatial coordinates to the coordinator drones for further action. Additionally, service drones equipped with an onboard \textit{fire detection} DNN model continuously monitor the landscape for signs of fire. If a fire is detected outside the known perimeter, the drone promptly notifies the base station, which updates the fire boundary.

\subsubsection{Coordinator Drones for Route Planning and Monitoring}
The number of coordinator drones (CDs) is comparatively fewer than service drones, as their primary function is to manage evacuation routes and coordinate drone-assisted rescue operations. They are optimally placed across the region to provide the maximum coverage.
When a service drone detects an individual in distress, it sends an assistance request to the nearest coordinator drone. The coordinator drone, upon receiving this request, processes the individual's current geolocation, cross-references it with pre-identified safe zones, and generates the most accessible and least hazardous escape route to reach there. 
The generated escape route is then transmitted to evacuees stranded on the ground, providing them with actionable guidance to reach a safe location. To enhance safety and reliability, coordinator drones continuously monitor the locations of evacuees, ensuring that the planned escape route remains viable. If fire spread threatens the designated route, alternative paths can be re-computed and updated dynamically.

\section{AeroResQ Workflow}\label{sec:workflow}

\begin{figure*}[t]
    \centering
    \includegraphics[width=0.9\linewidth]{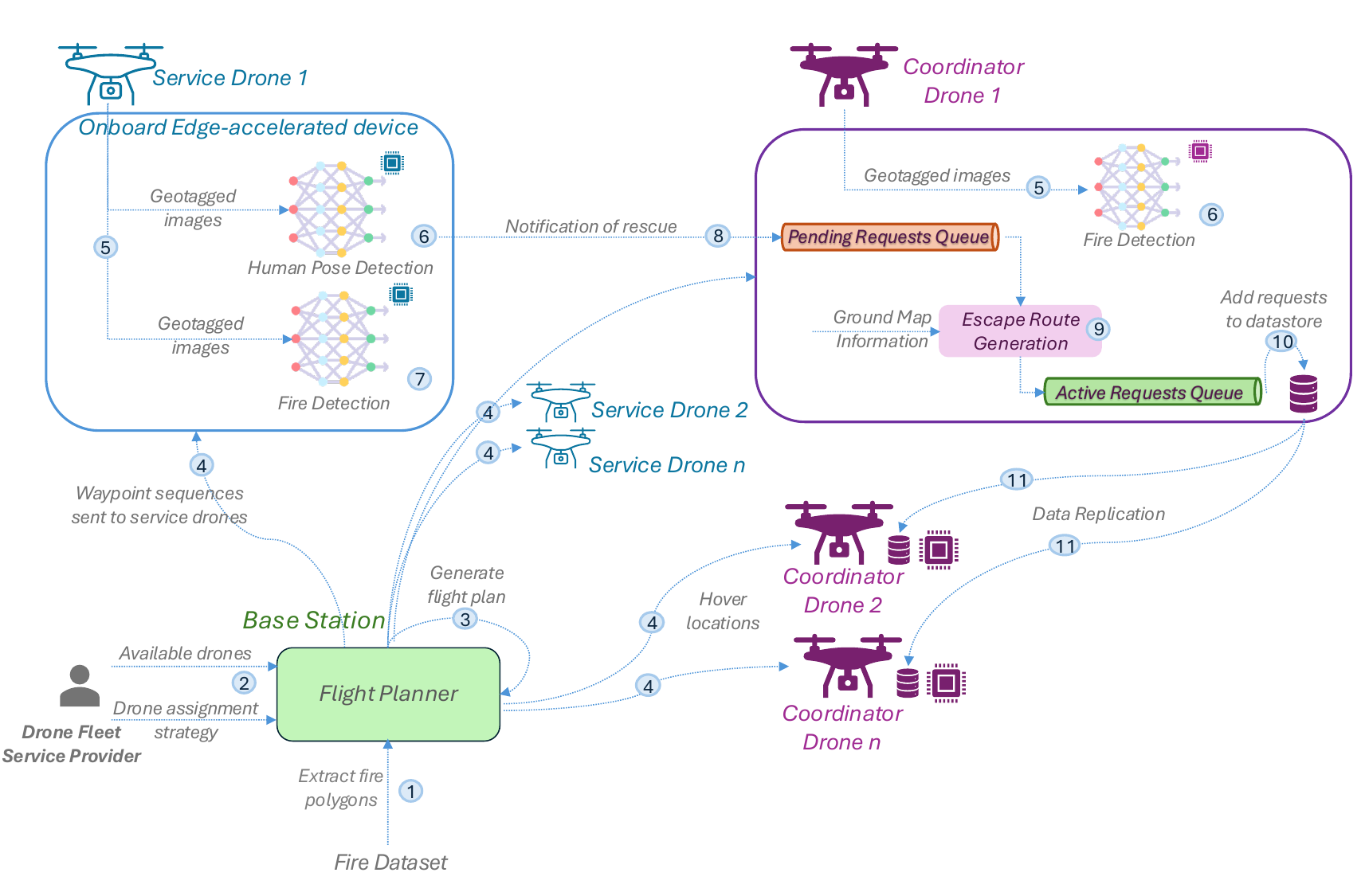}
    \caption{Workflow and Execution Sequence for AeroResQ.}
    \label{fig:arch}
\end{figure*} 

In this section, we discuss the system assumptions for the system requirements discussed above, provide a high-level workflow and then detail out the architecture of AeroResQ. 

\subsection{System Design Assumptions}
All UAVs are initially deployed from a base station on the ground, to which they may return for battery recharging if necessary. Each drone is equipped with at least an onboard camera and sensors for real-time observation, GPS for localization, infrared (IR) sensors and edge computing for onboard processing. Specifically, service drones are fitted with a forward-facing camera for detecting fires along their trajectory and a downward-facing camera for identifying evacuees. Coordinator drones, in contrast, may have a single camera with tilt flexibility, allowing it to face forward during movement for fire detection and to capture a bird’s-eye view when hovering.

Coordinator drones are equipped with powerful edge-
accelerated computing devices, such as the NVIDIA Jetson Orin AGX. Additionally, they host a distributed and lightweight data store onboard.
In contrast, service drones have onboard compute capabilities that are less powerful compared to coordinator devices, such as NVIDIA Jetson Orin Nano, as their primary role is localized monitoring. Once deployed, both coordinator and service drones maintain constant hovering and flying altitudes, respectively. For simplicity, we assume homogeneity within each category of drones. To enable autonomous operation, DNN models are pre-loaded, and all necessary software libraries are pre-installed. 

In \textit{AeroResQ}, base stations play a vital role in supporting UAV operations.  Beyond their logistical function, base stations are equipped with computing resources that act as a fallback option in scenarios where coordinator drones are unavailable. This ensures that essential processing tasks, such as route guidance and situational awareness, can continue without disruption. For communication, UAVs connect to a local WiFi hotspot or a cellular network via a full-duplex channel. While inter-drone communication occurs over WiFi, communication with the base station relies on cellular networks due to the extended distance range requirements. For simplicity, we assume that a battery swap occurs whenever the drone visits the base station, and the swap time is negligible.

\subsection{Workflow Description} 
The flow diagram in Fig.~\ref{fig:arch} represents the operational workflow of \textit{AeroResQ}, a dual-layer drone-based system designed for wildfire management and firefighter assistance. The process begins with the BS extracting fire perimeter data from the available fire polygon spatial information, such as NIFC~\cite{NIFC_Wildfire_Data}, which provides critical information on wildfire regions through satellite imagery and other geospatial sources. Based on the fire polygons and spatial partitioning strategy (\S~\ref{sec:spatial-partitioning}), the \textit{Flight Planner} at the BS generates flight plans, determining the optimal allocation of drones to meet the application requirements. The DFSP identifies available drones and assigns them roles: (1) service drones for localized surveillance, and (2) coordinator drones for high-level oversight. The BS then transmits the generated waypoint sequences to the SDs that depart from BS, enabling them to follow designated flight paths along the fire perimeter, while CDs are assigned hover locations at higher altitudes to maintain a broader field of view.

As the SDs navigate their assigned paths, they utilize onboard cameras to capture images of the fire-affected regions. These images are geotagged and processed in real-time using DNN models deployed on the onboard edge-accelerated computing devices. Specifically, SDs run human pose detection models to identify stranded individuals who require rescue, as this task requires closer proximity to accurately detect individuals. In contrast, fire detection can be performed from a higher altitude, allowing both SDs and CDs to execute fire detection models, and send the location of fire detection to the BS which dynamically update the fire perimeter.
If an SD detects a person in distress, it sends a notification of rescue to the nearest CD, which then takes over the rescue coordination process. The CD receives and processes the evacuation request, adding it to a \textit{Pending Requests Queue}, while simultaneously analyzing updated geotagged images to validate fire locations and ensure safe escape routes.

The next phase involves \textit{escape route generation}, a critical function performed by the CDs using ground map information and path planning algorithms. 
Once a suitable path is determined, the information is sent to the \textit{Active Requests Queue}, where it remains accessible for rescue teams to track the progression of evacuees along the path. 
To ensure system resilience, the CD hosts a lightweight data store that logs all active requests. This data is continuously replicated across multiple CDs, providing fault tolerance in case of drone failure due to extreme fire conditions or uncertain conditions. The data replication mechanism ensures that even if a CD is lost, other CDs in the fleet can seamlessly take over ongoing rescue operations without disrupting firefighter guidance.

In summary, AeroResQ enhances the efficiency of wildfire response efforts through the integration of autonomous UAV operations, onboard AI inferencing, dynamic flight planning, and resilient data management.

\section{Collaborative Escape Route Planning of Evacuation Requests}\label{sec:collaborative-planning}

In this section, we discuss the various strategies and algorithms that enable the system to collaboratively generate escape routes for the firefighters.

\subsection{Service drone assignment strategy}\label{sec:spatial-partitioning}
To establish the fire-affected region, the fire polygon is extracted as a sequence of waypoints represented by latitude-longitude pairs, forming the perimeter of the fire. We enhance spatial accuracy by constructing a continuous boundary representation by connecting consecutive waypoints with straight-line segments, forming a spline approximation of the fire perimeter. We further refine this polygonal representation by interpolating additional waypoints along each segment, ensuring that the distance between consecutive waypoints does not exceed $10$ meters. This gives better spatial detection and uniform coverage.

Once the fire polygon is fully discretized into waypoints, we allocate them to service drones for surveillance. For this, we randomly sample $|S|$ waypoints from the fire polygon, where $|S|$ represents the total number of available service drones. These sampled waypoints serve as the initial cluster centroids. We then apply a modified K-Means clustering algorithm, executing a single iteration to form waypoint clusters based on a chosen distance metric (e.g., Haversine distance for geospatial data). Each cluster of waypoints is subsequently assigned to a corresponding service drone, ensuring an efficient distribution of the surveillance task across the fleet. This approach helps optimize coverage while maintaining computational efficiency, allowing the service drones to monitor the fire perimeter effectively.

When assigning waypoint clusters to service drones, the base station simultaneously evaluates the feasibility of each drone completing its assigned surveillance task within its available onboard energy. This feasibility check ensures that the assigned cluster of waypoints can be fully traversed by the respective service drone without exhausting its battery. This considers factors such as the total distance to be covered, expected flight duration, and the energy consumption model of the drone. If the base station determines that the given number of service drones is insufficient to fully monitor the fire polygon due to energy constraints, an exception is triggered. In such cases, additional service drones will need to be deployed to handle the workload. This adaptive resource allocation mechanism ensures uninterrupted and efficient surveillance, preventing coverage gaps in critical fire-affected areas.

\subsection{Coordinator drone placement strategy}
The placement of coordinator drones is determined at the start of the mission to ensure optimal coverage of the fire-affected region. Since coordinator drones are responsible for providing a high-level overview of the area, their hover points must be strategically selected to maximize coverage while minimizing redundancy. To achieve this, we position the coordinator drones far apart across the fire polygon, effectively covering the entire fire-affected zone.

This hover point selection follows a systematic approach. Initially, a large number of candidate GPS coordinates (100s) are randomly sampled from the fire region. These serve as potential hover locations. The first hover location is selected arbitrarily from this set. Next, we identify the waypoint that is farthest from the first hover location among the remaining candidates and designate it as the second hover location. Subsequently, the third hover location is chosen as the waypoint that is farthest from both the first and second hover locations. This iterative process continues, with each new hover point being selected based on its maximum distance from all previously chosen hover locations. The process terminates once the number of hover locations equals the number of available coordinator drones.

By employing this \textit{farthest-first selection strategy}, the coordinator drones are distributed in a manner that maximizes spatial separation, thereby ensuring broad coverage of the fire region.

\subsection{Evacuation Request Description}\label{sec:evacuation-request-creation}
Once the drones are deployed, the \textit{coordinator drones} assume their designated hover positions, while the \textit{service drones} begin traversing the predefined waypoints assigned by the flight planner. During their flight, service drones continuously capture live video feeds whose image frames are analyzed in real time by a body pose estimation model to detect the presence of individuals in distress.

Let $\mathcal{R}$ represent a request that is generated when a service drone detects a stranded individual. Each evacuation request $\mathcal{R}_n$ is characterized by the following tuple: $\langle rID_n, t^d_n, t^e_n,$ $\lambda^d_n, [\lambda^s_n],$ $sID_n, cID^e_n,$ $cID^p_n,$ $[\lambda^{esc}_n], \lambda^f_n, t^f_n, ind_n,$ $len_n, status_n \rangle$. Here, $rID_n$ is a Universally Unique Identifier (UUID) assigned to the request, $t^d_n$ is the timestamp at which the service drone detected the individual, $t^e_n$ is the time at which the request was logged into the onboard data store, and
$\lambda^d_n$ is the geographical coordinates (latitude, longitude) of the detection. Since the drone employs a downward-facing camera for firefighter detection, the location of the drone at the moment of detection is assumed to be the same as that of the detected person.
$[\lambda^s_n]$ is a list of predefined safe locations near the fire-affected areas where evacuees can be safely relocated. 
$sID_n$ identifies the service drone that detected the individual and initiated the request.
$cID^e_n$ refers to the coordinator drone that initially receives the request. This is typically the spatially closest coordinator drone to the service drone $sID_n$ at the time of detection.
$cID^p_n$ denotes the coordinator drone responsible for processing the request. It is possible that the receiving coordinator drone ($cID^e_n$) may be different from the drone that ultimately processes the request ($cID^p_n$).

$[\lambda^{esc}_n]$ is the set of waypoints generated by the escape route planning algorithm (\S~\ref{sec:escape-route-generation}). These form a macro-level path that guides individuals toward safe locations.
$\lambda^f_n$ represents the last known location of the person for whom the request was raised, and $t^f_n$ records the most recent timestamp at which this location was updated. These parameters allow continuous tracking of the individual’s movement until they reach safety.
$ind_n$ denotes the index of the next waypoint the individual should proceed to, aiding in the guided walk strategy (\S~\ref{sec:guided-walk-strategy}).
$len_n$ specifies the total length (in meters) of the generated escape route.
$status_n$ represents the current state of the request, which can take one of three values:
\texttt{PENDING}: The request is newly received by the coordinator drone and is awaiting processing.
\texttt{ACTIVE}: An escape route has been successfully generated for the request, and the individual is in the process of evacuation.
\texttt{PROCESSED}: The individual has reached a designated safe location, and the request is considered resolved.

\subsection{Escape Route Generation}\label{sec:escape-route-generation} 
To generate an escape route for an evacuation request, we begin by obtaining the person's current location, denoted as $\lambda^d_n$. Based on a predefined set of safe locations within a region, we compute a sequence of waypoints that guide the individual to safety. This route is determined by minimizing both the travel distance and the elevation gain while avoiding hazardous fire zones. 

\noindent \textbf{Ground Map Generation:} The base map of the fire-affected region is represented as a graph derived from OpenStreetMap (OSM)~\cite{openstreetmap}. The graph is created using OSM's API by specifying the bounding box coordinates of the geofenced area. OSM provides flexible options for constructing the graph, allowing users to specify whether they want to include pathways suitable for driving, walking, or all modes of transport. In this graph representation, nodes correspond to road intersections or junctions, while edges denote pathways connecting these nodes. OSM assigns a distance (in meters) to each edge, which is stored as an edge attribute in the graph.

\noindent \textbf{Graph Pruning:} To account for the spread of fire, we refine the generated graph by removing nodes and edges that intersect with the fire perimeter. This is achieved using the \texttt{shapely}~\cite{shapely} Python library for spatial analysis and geometric transformations. If a node is located within the fire perimeter, it is removed from the graph along with all its associated edges. This pruning step is performed once at the start of the mission so that it is not accounted during route computation to ensure real-time performance during the evacuation process. 

Once the fire-pruned graph is obtained, we augment it with elevation data using the Elevation API from Google Maps~\cite{googleelevation}. To determine elevation gain along the paths, we interpolate $n$ intermediate GPS points on each edge and assign corresponding elevation values to them. These elevation metrics play a crucial role in computing an optimal escape route. The post processing of ground map graph is described in Algorithm~\ref{algo:graph-postprocessing}. Once processed, these pruned graphs are loaded to the coordinator drones, which is a one-time process and does not include any overheads during runtime. These graphs can be updated later based on the fire spread. 

\begin{algorithm}[!t]
\caption{Ground Map | Graph Post-processing}
\label{algo:graph-postprocessing}
\begin{algorithmic}[1]
\State \textbf{Step 1: Graph Pruning}
\For{each node $v \in V$}
    \If{$v$ inside fire polygon}
        \State Remove $v$ and its edges from $G$
    \EndIf
\EndFor

\State \textbf{Step 2: Augment Graph with Elevation Data}
\For{each edge $e = (u, v) \in E$}
    \State Sample intermediate points $\{p_1, p_2, \dots, p_m\}$ along $e$
    \State Query Google Elevation API for elevation at each $p_i$
    \State Compute elevation gain $\Delta h(e) = \max(0, h(v) - h(u))$
\EndFor
\end{algorithmic}
\end{algorithm}

\begin{algorithm}[!t]
\caption{Escape Route Generation Algorithm using weighted A* Search}
\label{algo:escape-route-gen}
\begin{algorithmic}[1]
\Require Pruned graph $G$, Current location $\lambda^d_n$, Safe Locations $S$, Weight Parameters $(\alpha, \beta)$
\Ensure Optimal Escape Route $P$
\State Compute origin node $n_o \gets \text{NearestNode}(G, \lambda^d_n)$
\For{each $s_i \in S$}
    \State $n_s \gets \text{NearestNode}(G, s_i)$
    \State $d(s_i)$ = \text{distance}$(n_o, n_s)$
\EndFor
\State Sort $S$ in increasing order of $h(s_i)$
\For{each $(n_s \in S)$}
    \State $w(e) = \alpha \cdot d(s_i) + \beta \cdot \Delta h(e)$\Comment{Calculate weighted cost function}
    \State $\mathcal{P} = \text{A*}(G, n_o, n_s, w)$ \Comment{Compute best path using weighted A*}
    \If{$\mathcal{P}$ exists}
        \State \textbf{return} $\mathcal{P}$
    \Else
        \State continue
    \EndIf
\EndFor
\State Notify base station
\end{algorithmic}
\end{algorithm}

\noindent \textbf{Route Generation using Weighted A* search:} We use the A* algorithm~\cite{hart1968formal} to identify the \textit{safest} and \textit{shortest} escape route over the graph (Algorithm~\ref{algo:escape-route-gen}). A* is a graph traversal and path-finding algorithm that efficiently finds the shortest path between nodes using a combination of actual cost from the start node $(g(n))$ and a heuristic estimate $(h(n))$ of the remaining cost to the goal. A* balances optimality and computational efficiency, making it well-suited for dynamic environments like wildfire evacuation, where minimizing both distance and elevation gain is crucial. 

\begin{figure}
    \centering
    \subfloat[$\beta = 0.1$]{
    \includegraphics[width=0.4\columnwidth]{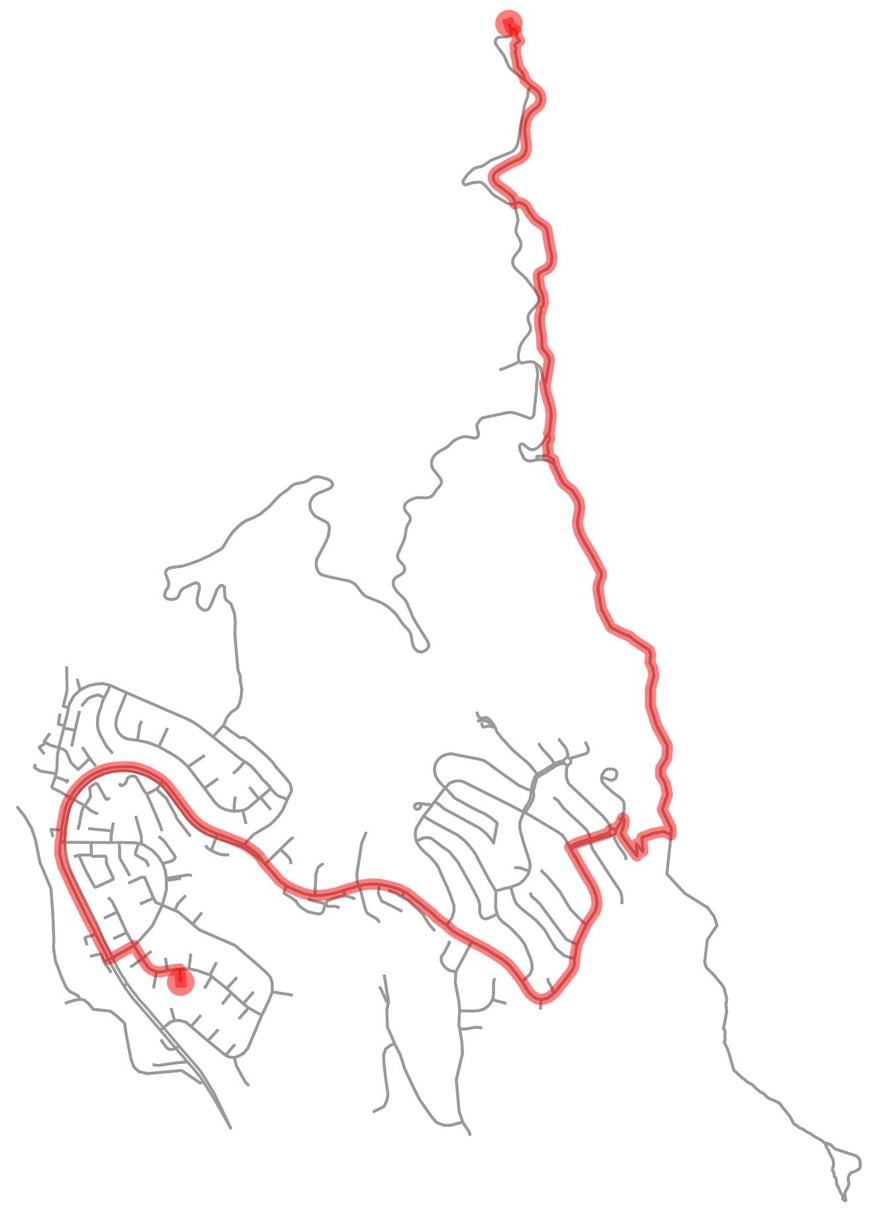}
    \label{fig:beta-0.1}
    }
    \subfloat[$\beta = 0.2$]{
    \includegraphics[width=0.4\columnwidth]{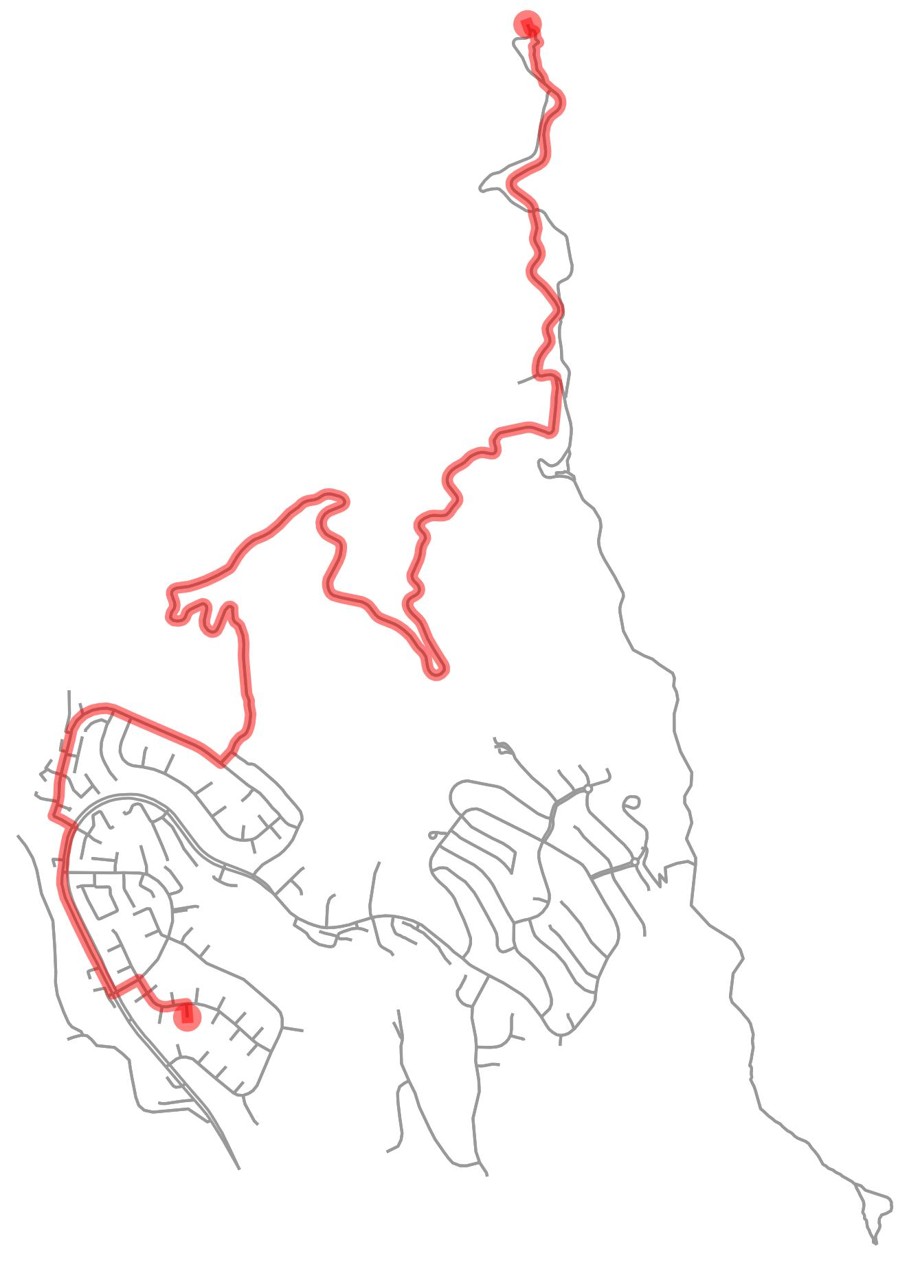}
    \label{fig:beta-0.2}
    }  
    \caption{Effect of elevation gain penalty on route generation by A* causes route to change.}
    \label{fig:beta-effect}
\end{figure}

The algorithm first loads the preprocessed graph having the elevation data. Given an individual's current location $\lambda^d_n$, we identify the nearest corresponding node in the graph, $N_o$, and map each safe location $s_i \in S$ to its nearest graph node $N_s$. 
These safe locations are then arranged in ascending order based on their edge distances. This ensures that the closest safe locations are prioritized when determining the optimal escape route. The algorithm then defines a weighted cost function,
    \[w(e) = \alpha \cdot d(s_i) + \beta \cdot \Delta h(e)\]
%
where $d(s_i)$ and $\Delta h(e)$ represent the distance and the elevation gain for the path, respectively, and $\alpha$ and $\beta$ are user-defined weights that balance distance minimization and elevation avoidance. Using this cost function, A* iteratively explores the shortest escape route for each sorted safe location, prioritizing paths that are both efficient and safe. If a valid path $P$ is found, the corresponding sequence of GPS waypoints is returned. Otherwise, the algorithm proceeds to search the path corresponding to the next nearest safe location until a feasible route is identified. If no path is found, the drone sends a notification to the base station to handle the request through other means of emergency response. This ensures that evacuees are guided along the safest possible route while minimizing travel effort and avoiding fire hazards effectively. Figure~\ref{fig:beta-effect} illustrates how varying the penalty on elevation gain ($\beta$) in A* search affects route generation while keeping $\alpha$ constant. In Fig.~\ref{fig:beta-0.1}, distance and elevation gain are weighted equally ($\alpha=1.0,\beta=0.1$), whereas Fig.~\ref{fig:beta-0.2} prioritizes minimizing elevation gain ($\alpha=1.0,\beta=0.2$).

Once the firefighters are notified of their designated escape routes, it is crucial to continuously monitor their movements to ensure they are following the intended paths and reach their safe locations successfully. This real-time monitoring is facilitated by coordinator drones, which periodically check the evacuee’s current position. In a real-world scenario, this can be achieved through a \textit{smartphone-based communication system}, where the firefighter’s mobile device interacts with the coordinator drone. The smartphone would receive the escape route details from the drone while simultaneously transmitting its real-time location back to the drone. This allows the drone to confirm that the firefighter is following the escape path. Once the firefighter reaches the safe location, the monitoring is terminated as the evacuation request has been completed.

\subsection{Request Replication using onboard datastore} 

\subsubsection{Choice of onboard distributed datastore technology} 
Various data stores cater to different workload needs, each with unique strengths and trade-offs. \textit{Redis}~\cite{redis_docs} offers high-performance, low-latency caching with built-in geospatial support but is memory-bound. \textit{Cassandra}~\cite{cassandra_docs} excels in write-heavy workloads with fault tolerance but requires external tools for spatial queries. \textit{InfluxDB}~\cite{influxdb_docs} is optimized for high-ingestion time-series data but restricts clustering to enterprise versions. \textit{MobilityDB}~\cite{mobilitydb_docs}, built on PostgreSQL/PostGIS, specializes in spatio-temporal workloads but has a complex setup. \textit{RiakKV}~\cite{riak_docs} ensures high availability and fault tolerance but lacks advanced query capabilities. \textit{IoTDB}~\cite{iotdb_docs} is tailored for IoT and sensor analytics, supporting distributed workloads but requiring plugins for geospatial indexing.

For AeroResQ, we choose \textit{IoTDB} as the primary datastore due to its high-throughput write operations, strong consistency, and low-latency reads, making it ideal for managing large-scale IoT sensor data in real-time wildfire monitoring scenarios. IoTDB internally uses Apache Ratis, an implementation of the Raft consensus protocol, to ensure strong consistency across distributed cluster deployments. While alternatives like InfluxDB offer similar time-series capabilities, IoTDB's focus on industrial IoT workloads, efficient compression, and better integration for distributed deployments aligns well with AeroResQ's requirements. Additionally, its scalability ensures that AeroResQ can handle increasing data volumes from UAV-based sensors and edge devices without compromising performance.

\subsubsection{Data replication using IoTDB}
Apache IoTDB follows a distributed architecture with three key components: Config Nodes, Data Nodes, and Seed Nodes. Config Nodes manage metadata, schema, and cluster configuration, ensuring consistency using a consensus protocol. They also coordinate the distribution of DataRegions across Data Nodes. Data Nodes handle storage, query execution, and data ingestion, with multiple DataRegions for time-series data and SchemaRegions for schema management. To enhance fault tolerance, both data and schema are replicated across nodes. Seed Nodes facilitate cluster initialization by helping new nodes join seamlessly. In our setup, each coordinator drone runs a Data Node and a Config Node, with one randomly chosen as the Seed Node. In our experiments, all the coordinator drones host one instance of a data node along with a config node. One of the coordinator drones is randomly chosen to be the seed node, and we set the data replication and schema replication as $3$.

\section{Resilience Algorithms}\label{sec:resilience-algorithms}

In this section, we introduce a set of algorithms that have the goal to increase the overall resilience of AeroResQ. With the support of these algorithms AeroResQ is able to ensure continuous and adaptive evacuation support in wildfire scenarios. UAVs operate in highly dynamic environments where hardware failures can compromise the reliability of escape route guidance. Without resilience mechanisms, failures such as drone malfunctions or lost connectivity could leave firefighters without updated evacuation paths, increasing their risk. To address these challenges, we discuss strategies to mitigate two critical failure scenarios: (1) failure of coordinator drones, which would result in evacuees being left unattended and unmonitored, and (2) failure of service drones, which would lead to unsupervised spatial regions lacking surveillance.

\subsection{Failure of Coordinator Drones}
To maintain an active record of coordinator drones operating during the mission, a heartbeat mechanism is implemented among the coordinator drones. This mechanism ensures continuous monitoring of drone availability and helps detect failures in real-time. One of the coordinator drones (CDs) is designated as the heartbeat client, while the remaining coordinator drones function as heartbeat servers. At regular intervals, the heartbeat client sends a request to all other coordinator drones, prompting them to respond with an acknowledgment. A drone is considered active as long as it continues to respond within an expected time window. However, if a drone fails to respond to multiple consecutive heartbeat requests over a prolonged period, it is flagged as lost. 

Each CD maintains a local list that records the current load on all coordinator drones in the system. This \textit{load} is defined as the sum of the escape route lengths assigned to each CD for all active evacuation requests. The heartbeat client plays a crucial role in synchronizing this information. As part of its periodic heartbeat message, the client includes the latest version of the load list, which contains the load of all CDs. When a heartbeat server responds to the client, it also sends back its current load. This bi-directional exchange ensures that with each heartbeat cycle, all coordinator drones remain updated on the distribution of active escape routes across the fleet. This decentralized awareness allows the system to make informed decisions in case of drone failures or dynamic workload rebalancing.

In the event of a CD failure, the system ensures seamless continuity by dynamically reassigning responsibilities. If the failed CD was acting as the heartbeat client, one of the remaining CDs is randomly selected to take over this role, allowing the \textit{load} list to continue updating without disruption. Additionally, the newly assigned heartbeat client initiates the redistribution of active evacuation requests previously managed by the failed CD. Since IoTDB employs a data replication mechanism, copies of the evacuation requests stored on the failed CD are also available on other CDs. To facilitate reassignment, a recovery algorithm queries the replicated data to identify requests where the processing coordinator drone ID ($cID^p$) corresponds to the failed CD. For each such request, the system runs a bin-packing algorithm, leveraging the local load list to redistribute requests among the remaining CDs in a way that maintains workload balance. This ensures that evacuation operations continue seamlessly, with minimal disruption.

Furthermore, upon detecting a CD failure, the base station recalculates the hover positions for the remaining CDs to maximize coverage of the fire region. The updated hover locations are then communicated to the CDs, which reposition themselves accordingly, ensuring effective surveillance and coordination despite the loss of a drone.

\subsection{Failure of Service Drones} 
Similar to the CD-CD heartbeat mechanism, we implement an SD-CD heartbeat protocol, where each service drone (SD) responds to periodic heartbeat requests sent by a coordinator drone (CD). If a CD fails to receive a response from any SD within a predefined time window, the system treats the unresponsive SD as potentially loss. In such cases, the base station is immediately notified, triggering a reallocation of surveillance responsibilities among the remaining service drones. The reassignment process follows the spatial partitioning strategy outlined in \S~\ref{sec:spatial-partitioning}. Given the reduced number of available service drones, the fire polygon is re-segmented to ensure continued coverage, albeit with adjusted workload distribution. This adaptive approach helps maintain the integrity of the surveillance mission, ensuring that fire monitoring and evacuee tracking remain uninterrupted despite drone failures.

\section{Architecture and Deployment}\label{sec:arch}
\subsection{Architecture}
The base station receives information about the fire polygon in geoJSON format from sources such as satellites or remote sensing authorities. The \textit{flight planner thread} partitions the fire polygon into clusters and assigns the sequence of waypoints to SDs. These assignments are then shared with the SDs using gRPC. Each SD runs a \textit{waypoint executor thread} that geotags the captured images with the current location of the SD. These images are passed on to the \textit{body pose estimation} (BP) and \textit{fire detection model} (FD) running onboard the edge accelerator of SD. The BP model is wrapped around a gRPC client, which sends a message to the CD when a human is detected. This message contains the SD id, time and location of detection. 

The \textit{Incoming Request Handler} thread running on the CD receives these messages and creates an \textit{evacuation request} while populating more fields and adds them to \textit{Pending Requests (PR) queue}. The \textit{escape route generation} thread polls each request from the PR queue, generates the escape route for the request, mark the request as \texttt{ACTIVE} and adds it to the \textit{Active Requests (AR) queue}. The \textit{add to datastore} thread polls active requests from the AR queue and sequentially inserts them into the data node hosted on the respective CD by creating a \textit{session} for connection. Finally, a \textit{firefighter thread} with guided walk strategy running on the CD queries the requests being addressed by the same CD, and updates the last known location and waypoint at regular intervals until the requests are marked as \texttt{PROCESSED}.

\subsection{Automated Deployment using Docker Compose}
We utilize Docker Compose to automate the setup of our emulated environment for experiments. Docker Compose is a powerful orchestration tool that efficiently manages multiple containers by streamlining operations such as starting, stopping, scaling, resource allocation (CPU, memory, and GPU), monitoring container status, and handling networking and service discovery. In our setup, Docker Compose enables the seamless deployment of containers representing the base station, service drones, and coordinator drones. Resource constraints for CPU, memory, and GPU are specified in the \texttt{docker-compose.yml} files for each service and are enforced by the container runtime at execution. Multiple instances of a service can be deployed using the \texttt{--scale} flag, enabling flexible deployment for simulating drone fleet expansions. By configuring CPU, memory, and GPU allocations, we ensure that each container accurately reflects the computational capabilities of different drone types. In order to facilitate service discovery, Compose automatically creates an isolated bridge network for inter-container communication, and each service is assigned a DNS entry based on its name. Compose also provides an option to assign custom hostnames that can be specified to uniquely identify containers. We use this to facilitate gRPC communication among the containers and track active containers throughout the experiments. To further enhance portability and efficiency, we pre-built Docker images for the base station, service drones, and coordinator drones, bundling all necessary code and dependencies for both linux/amd64 and linux/arm64 platforms.

\section{Experiments} \label{sec:results}

\subsection{Setup}\label{sec:exp:setup}
We use Docker Compose to set up the emulation environment, which launches the required number of containers corresponding to service drones, coordinator drones, and the base station. These containers are interconnected via a \textit{bridged network} and run on a host server equipped with an AMD EPYC 7352 CPU with $64$~vCPUs and $80$~GB RAM, and two NVIDIA A10 GPUs with $9216$ CUDA cores, $288$ Tensor cores, and $24$~GB of GPU memory. The server operates on Rocky Linux release 8.10 (Green Obsidian) with CUDA Version 12.8. Resource allocation per container is as follows: each service drone is assigned $1$~vCPU and $1$~GB RAM, each coordinator drone receives $2$~vCPUs and $4$~GB RAM, and the base station is allocated $8$~vCPUs and $8$~GB RAM. For each CD, IoTDB is allocated $3$~GB out of $4$~GB memory. To balance computational load, the service and coordinator drone containers are equally divided between the two GPU. 

We evaluate collaborative route planning and resilience algorithms using three fleet sizes: \texttt{SMALL} ($3$~CD, $10$~SD), \texttt{MEDIUM} ($5$~CD, $20$~SD), and \texttt{LARGE} ($8$~CD, $40$~SD), ensuring feasibility within the host server's capabilities. For fire detection, we train the YOLOv11 (medium) model on two fire datasets: FlameVision~\cite{Jafar2023FlameVision} using $3150$ training images and the Fire dataset on Roboflow~\cite{fire-data-annotations_dataset} using $2621$ training images for 200 epochs, which we call as YOLOv11-1 and YOLOv11-2 respectively. To evaluate their accuracy, we test the resulting models on the remaining unseen test datasets not used for training and summarize the results in Table~\ref{tab:fire-accuracy}.
Since YOLOv11-2 model achieve better accuracy on the cross-datasets, we select it for our experiments. For body pose detection, we use NVIDIA's \texttt{trt\_pose} model~\cite{NVIDIA_TRTPose}, which is coupled with an SVM classifier~\cite{raj2023ocularone} to classify different human poses.

\begin{table}[!t]
    \centering
    \footnotesize
    \begin{tabular}{l|l|r}
        \hline
        \textbf{Fire Detection Model} & \textbf{Testing Data} & \textbf{Accuracy} \\ 
        \hline
        \multirow{3}{*}{ YOLOv11-1} 
            & Fire data (Roboflow) & 96\% \\ 
            & FlameVision & 68\% \\ 
        \hline
        \multirow{3}{*}{YOLOv11-2} 
            & Fire data (Roboflow) & 64\% \\ 
            & FlameVision & 80\% \\ 
        \hline
    \end{tabular}
    \caption{Accuracy of fire detection models.}
    \label{tab:fire-accuracy}
\end{table}

\subsubsection{Workloads} 

\begin{table}[!t]
    \centering
    \footnotesize
    \setlength{\tabcolsep}{2.5pt}
    \begin{tabular}{l|r|r|r}
        \hline
        \textbf{Fire} & \textbf{\# of Polygons} & \textbf{Area of Fire (in acres)} & \textbf{\# of Safe Locations} \\
        \hline
        Kenneth  & 1  & 998  & 7 \\
        Hughes   & 3  & 10425 &  68\\
        Eaton    & 3  & 14021 & 13\\
        Palisades & 4  & 23448 & 10\\
        \hline
    \end{tabular}
    \caption{Fire incidents with number of polygons and affected area.}
    \label{tab:fire_data}
\end{table}

\begin{figure}[!t]
    \centering
    \subfloat[Kenneth ($\approx1K$ acres burned)]{%
    \includegraphics[width=0.54\columnwidth]{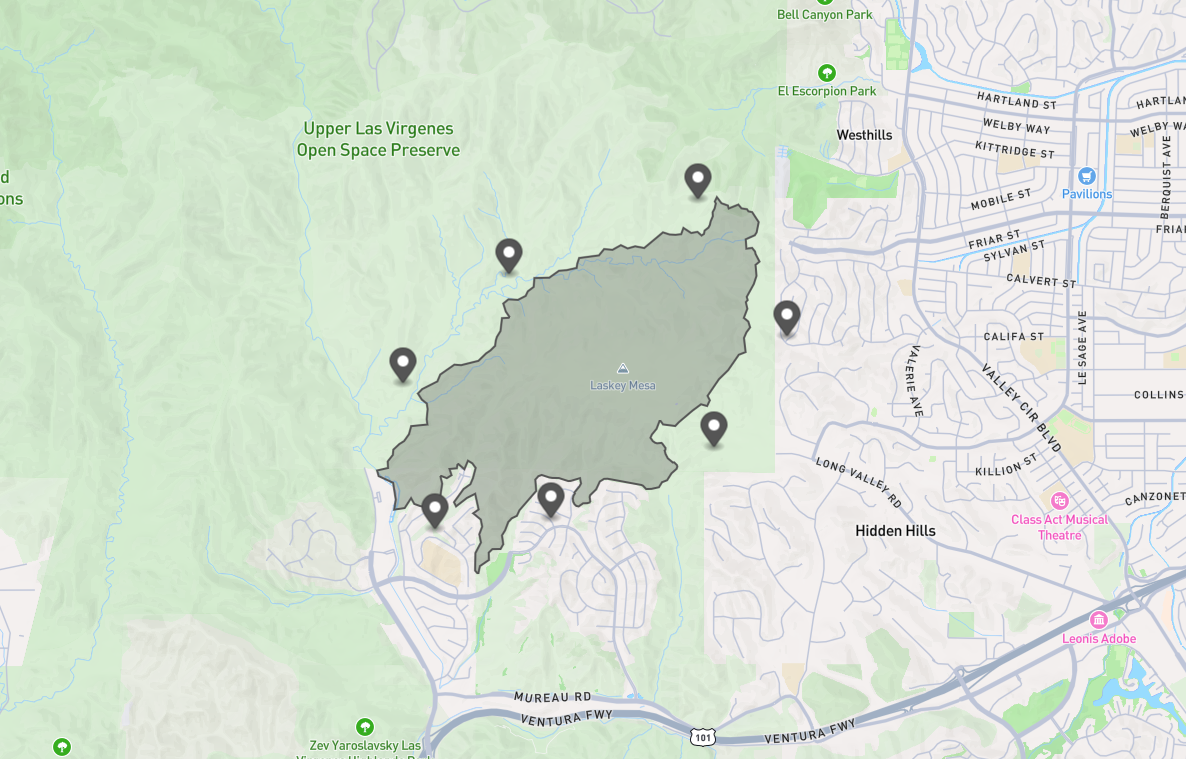}
    \label{fig:kenneth-safe}
    }%
    \subfloat[Palisades ($\approx23K$ acres burned)]{%
    \includegraphics[width=0.44\columnwidth]{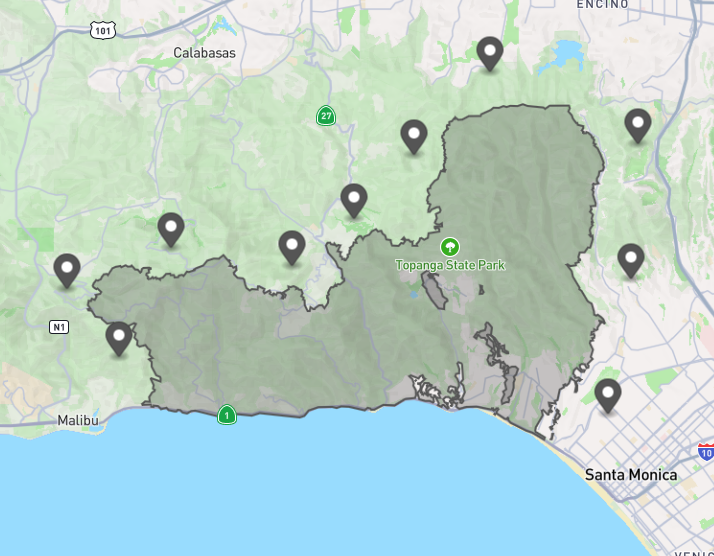}
    \label{fig:palisades-safe}
    }%
    \caption{Safe locations around fire polygons.}
    \label{fig:safe-locations}
\end{figure}

The fire polygons are obtained from the \textit{WFIGS 2025 Interagency Fire Perimeters to Date}~\cite{NIFC_Wildfire_Data} dataset in geoJSON format, which is maintained by the National Interagency Fire Center (NIFC). From this dataset, the \textit{coordinates} field within \textit{features} is processed into a sequence of waypoints represented as latitude-longitude pairs. Specifically, we extract wildfire data for Southern California in January 2025, focusing on the Kenneth, Hughes, Eaton, and Palisades fires (Table~\ref{tab:fire_data}). The safe locations around the fire polygon are manually marked using geojson.io with a visualisation shown in Fig.~\ref{fig:safe-locations}. 

We create a total of $12$ workloads that corresponds to fire region and drone fleet combination. \texttt{SMALL} (13) and \texttt{MEDIUM} (25) fleet sizes were not sufficient for Palisades, and \texttt{SMALL} (13) was not sufficient for Eaton fire, because of their large fire area sizes. The service drones operate with two datasets: one containing fire images, and the other including both fire and human images. Each SD travels at a speed of $5m/s$, following flight paths generated with a waypoint granularity of $10m$. Upon reaching each waypoint, an SD publishes an image, maintaining a frequency of one frame every $2$ seconds. Initially, the probability of selecting an image containing both fire and human presence is set at $0.8$, gradually decreasing to $0.2$ by the end of the experiment. This models a real-world scenario where the number of evacuees requiring assistance declines over time. The heartbeats are sent at an interval of $30$~seconds. 
Each CD runs a thread that implements a guided walk simulator (\ref{sec:guided-walk-strategy}) to update the locations of evacuees at a regular interval of $10$~seconds, leveraging IoTDB updates for precise localization. The number of evacuees for a given workload is determined solely by the number of service drones available, rather than the size of the fire.

\subsection{Results} 

\subsubsection{Overhead of the AeroResQ Architecture}

\begin{figure*}[t]
    \centering
    \subfloat[Kenneth]{
    \includegraphics[width=0.4\linewidth]{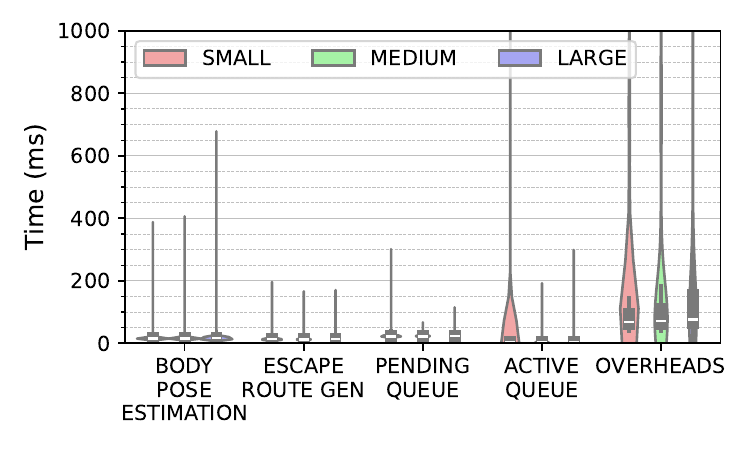}
    \label{fig:kenneth-30}
    }
    \subfloat[Hughes]{
    \includegraphics[width=0.4\linewidth]{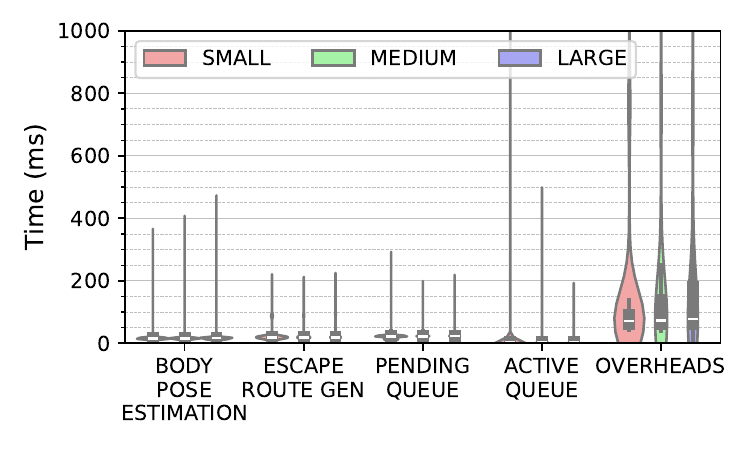}
    \label{fig:hughes-30}
    }\\
    \subfloat[Eaton]{
    \includegraphics[width=0.4\linewidth]{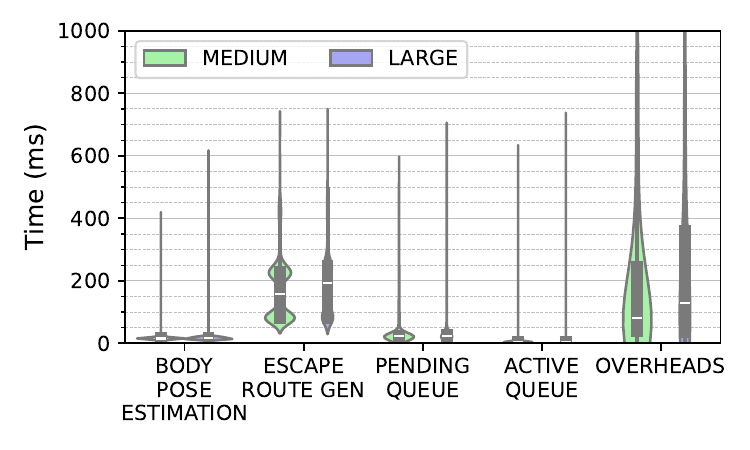}
    \label{fig:eaton-30}
    }
    \subfloat[Palisades]{
    \includegraphics[width=0.4\linewidth]{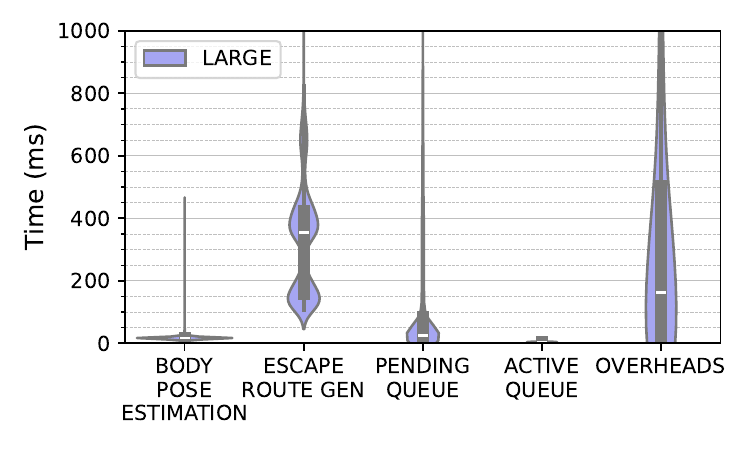}
    \label{fig:palisades-30}
    }
    \caption{Latency for different components of evacuation request creation for different drone fleet sizes (Small:13, Medium:25, Large:48), 30 minutes surveillance duration}
    \label{fig:aeroresq-overhead}
\end{figure*}

For a $30$ minutes surveillance duration by service drones, we report that the overhead of the AeroResQ architecture is minimal, with a nominal end-to-end latency of $\leq500$~ms per evacuation request.
For each request generated by a service drone, we measure the overheads introduced by AeroResQ, focusing on the end-to-end latency from request creation at the service drone to its insertion into the datastore on the coordinator drone. This includes the time taken for body pose estimation and escape route generation. In Fig.~\ref{fig:aeroresq-overhead}, we report the Body Pose Estimation (BPE) inference time, escape route generation time and overheads for \texttt{SMALL}, \texttt{MEDIUM} and \texttt{LARGE} fleet sizes. We observe that within a specific fire region i.e., Kenneth fire (Fig.~\ref{fig:kenneth-30}),, the processing times for each component remain consistent across all service drones in the fleet. BPE, executed on the GPU, achieves a median inference time of $20$~ms, indicating that the GPU efficiently handles inferencing requests from multiple service drones with minimal variability. Escape route generation reports a median latency of $40$~ms, contributing to a combined processing time of approximately $50$~ms. Additional overheads, such as queuing delays in the PR and AR queues, exhibit a median latency of $\leq50$~ms. Overall, our platform achieves a nominal end-to-end latency of $\leq500$~ms per request, demonstrating its suitability for real-time deployment in onboard AI-driven wildfire response applications even for fleets with 48 drones.

\subsubsection{Evacuation Requests in IoTDB }

\begin{figure}[t]
    \centering
    \includegraphics[width=0.45\linewidth]{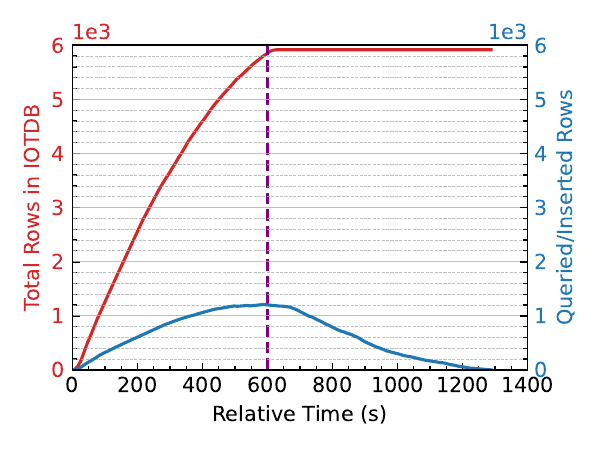}
    \caption{Variation of \# of updated rows with respect to experiment timeline. The dotted vertical line shows the end of this service drone's surveillance.}
    \label{fig:db-update-guided-walk}
\end{figure}

Since the guided walk simulator continuously updates the last known waypoint and timestamp of evacuation requests, the evacuee proxy thread periodically queries active requests from \textit{IoTDB}. To analyze the correctness of request processing, we focus on one specific coordinator drone. The X-axis represents the relative time (in seconds) from the start of the experiment, while the left Y-axis denotes the total number of rows in \textit{IoTDB}, and the right Y-axis indicates the number of updated rows.

As the experiment begins, we observe a steady increase in the total number of rows in \textit{IoTDB}, corresponding to newly created evacuation requests. Simultaneously, the number of rows requiring updates, reflecting requests with location modifications, also increases. When the experiment concludes at $600$~seconds, the total number of rows in \textit{IoTDB} stabilizes, as no additional requests are generated by the service drones. Consequently, the number of updates gradually decreases over time, simulating the evacuees walking towards their safe location, and by $1250$~seconds, all requests are marked as \texttt{PROCESSED}.

\subsubsection{Scalability of AeroResQ}

\begin{figure*}
    \centering
    \subfloat[60 minutes]{
    \includegraphics[width=0.4\linewidth]{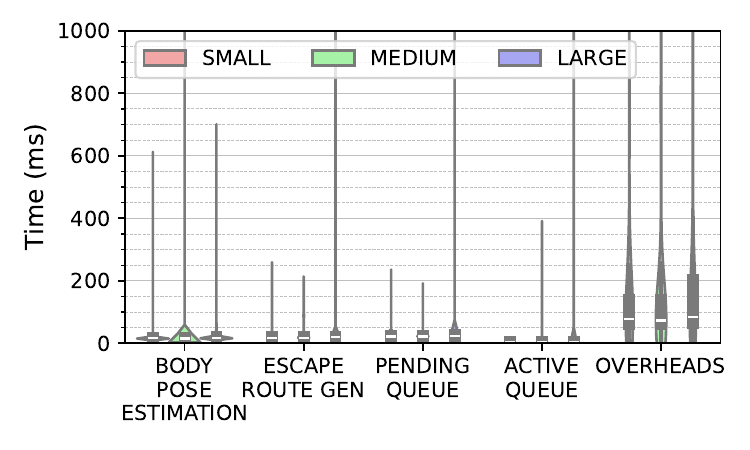}
    \label{fig:hughes-60}
    }
    \subfloat[180 minutes]{
    \includegraphics[width=0.4\linewidth]{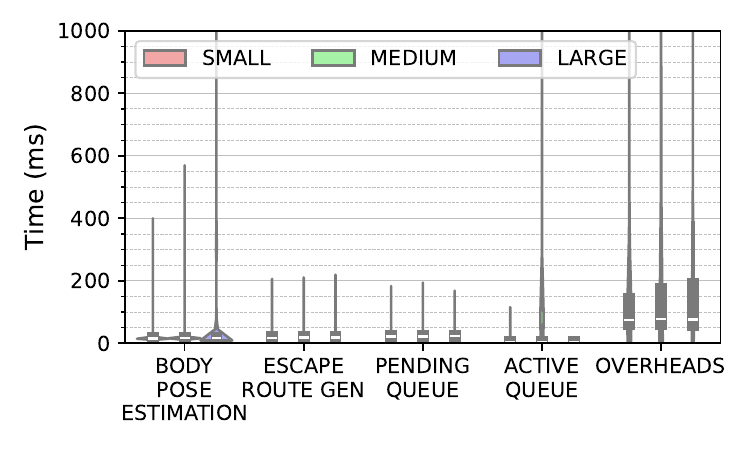}
    \label{fig:hughes-180}
    }
    \caption{Latency for different components of evacuation request creation for different drone fleet sizes (Small:13, Medium:25, Large:48) and different durations of surveillance for Hughes Fire region}
    \label{fig:hughes-variable-duration}
\end{figure*}

Fig.~\ref{fig:aeroresq-overhead} presents the scalability analysis of the AeroResQ platform across different fire regions: Kenneth (Fig.~\ref{fig:kenneth-30}), Hughes (Fig.~\ref{fig:hughes-30}), Eaton (Fig.~\ref{fig:eaton-30}) and Palisades (Fig.~\ref{fig:palisades-30}), and different fleet sizes: \texttt{SMALL}, \texttt{MEDIUM} and \texttt{LARGE} for a surveillance duration of $30$ minutes. We observe that there is a negligible impact on BPE inference times and AR queue time throughout these scenarios. Interestingly, we see that the time to generate escape routes increases with an increase in the area of the fire region. 
The variation in escape route generation time can be attributed to the graph size used in A* search. The weighted A* algorithm, used in our experiments, has a time complexity of $O(b^d)$, where b is the branching factor and d is the depth of the search. As the graph size increases (i.e., more nodes and edges representing escape routes in complex wildfire regions), the search space expands, leading to increased processing time. In the figure, scenarios with larger ground map areas, i.e., graph size, such as Eaton and Palisades, result in higher route generation times due to a greater number of waypoints to evaluate.
Moreover, the increased variance in execution time for Eaton and Palisades suggests that some routes require significantly longer computations, possibly due to denser fire regions or more complex terrain constraints, leading to increased node expansions in A*. In contrast, Kenneth shows minimal variance, indicating a relatively simple escape route with fewer diversions, resulting in near-constant computation time. 

As a result, tasks experience longer wait times in the PR queue before being scheduled for escape route generation. This increased queuing delay contributes to higher overheads, which is most evident in the Palisades region, showing the longest PR wait time and the highest overall overhead among all fire regions. Nevertheless, across all configurations, AeroResQ remains lightweight, achieving a median end-to-end latency of $\leq500$~ms, thereby demonstrating its scalability and efficiency under varying workloads. To further evaluate AeroResQ scalability across longer surveillance durations, we conduct experiments of $1$ and $3$ hours for the Hughes fire region, shown in Fig.~\ref{fig:hughes-variable-duration}. The trends observed are consistent with those in Fig.~\ref{fig:hughes-30}, reaffirming that the AeroResQ architecture maintains its scalability and efficiency even under prolonged operational conditions.

\subsubsection{Adaptation to Coordinator Drones Failures}

\begin{figure}[t]
    \centering
    \includegraphics[width=0.6\linewidth]{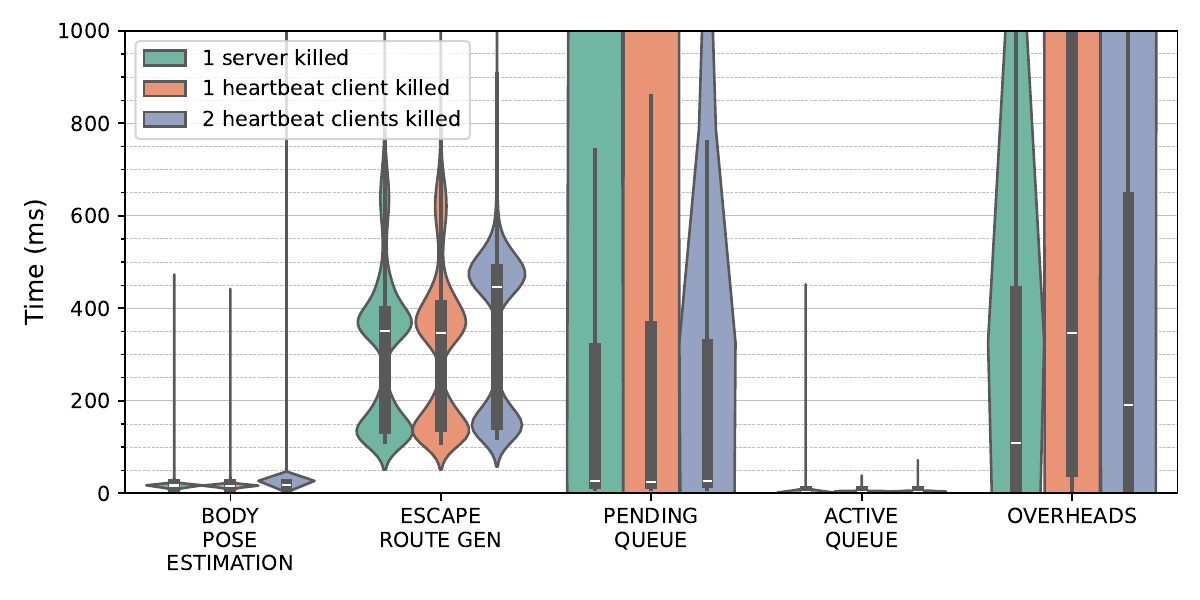}
    \caption{Latency for different components of evacuation request creation for LARGE drone fleet size in Palisades fire region, 30 minutes surveillance duration and different CD failure scenarios}
    \label{fig:aeroresq-resilience}
\end{figure}

\begin{figure}
    \centering
    \includegraphics[width=0.5\linewidth]{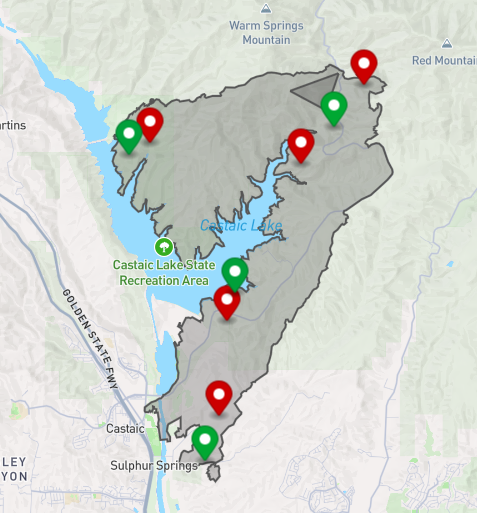}
    \caption{Updated hover locations of four coordinator drones (green) after failure of the fifth. Previous positions are shown in red.}
    \label{fig:cd-resilience}
\end{figure}

We evaluate the robustness of AeroResQ under coordinator drone (CD) failures, using three failure scenarios on a \texttt{LARGE} fleet configuration comprising $8$ CDs deployed over the Palisades fire region: (A) single heartbeat client failure, (B) single heartbeat server failure, and (C) multiple heartbeat client failures. Each experiment lasted $30$~minutes, with the first failure triggered approximately $4$~minutes after initiation, and in scenario (C), the second failure occurred around $6$~minutes after the first. As shown in Fig.~\ref{fig:aeroresq-resilience}, the system maintains stable performance across all failure scenarios. In Scenario A, $757$ requests were successfully load-balanced within $1018$~ms, while $673$ requests were handled within $1553$~ms in Scenario B, reflecting the additional overhead of restarting a new client. Since each CD handles a varying number of requests being received by SDs depending on its initial hover location, the observed overheads are influenced by the workload of the failed CD. In scenario~(C), only $86$ requests were processed in $184$~ms following the first failure and $386$ requests in $888$~ms after the second, which reflects in the overheads accordingly. These requests were originally being handled by the failed drone. As the median end-to-end latency is $\leq500$ ms for all scenarios, this demonstrates that the load-balancing mechanism is lightweight and ensures minimal disruption. Additionally, Fig.~\ref{fig:cd-resilience} visualizes the pre- and post-failure coordinator drone positions for \texttt{HUGHES} fire region, where the red markers (5) represent the original hover locations, and the green markers (4) denote the updated positions of the remaining CDs after reallocation, after the fifth drone in the top right fails.

\section{Related Work}\label{sec:related}
 In this section, we review existing works on UAV-based wildfire operations, resilient multi-UAV coordination, and escape route planning, highlighting their contributions and identifying gaps that \textit{AeroResQ} aims to address.
 
\subsection{Use of Drones for WildFire Management}
Several studies have investigated the deployment of UAVs for wildfire management, focusing on tasks such as fire detection, monitoring, and emergency response~\cite{drones8050203}. Many approaches utilize drones equipped with thermal cameras and deep learning models to detect fire hotspots in real time~\cite{rs15071821,8845309}. Some efforts~\cite{10681400} integrate spatio-temporal environmental data captured by UAVs to enhance wildfire risk estimation and support proactive mission planning. Others~\cite{9953997} use UAVs primarily for data collection, enabling subsequent fire detection, segmentation, or modeling through data-driven techniques. 

The DOME system~\cite{10.1145/3576841.3585929} proposes a drone-assisted monitoring framework that coordinates multiple heterogeneous UAVs to gather real-time situational data during evolving emergency events, with its effectiveness evaluated in simulated prescribed burns. Additionally, UAVs are utilized for mapping wildfire-prone regions to facilitate preventive measures~\cite{f14081601} and for post-disaster surveillance, aiding in damage assessment~\cite{drones3020043}. In contrast, \textit{AeroResQ} goes beyond fire detection and monitoring by integrating a dynamic escape route planning framework, actively assisting evacuees in navigating toward safe zones.

\subsection{Escape Route Generation}
Few studies have explored UAV-based path-planning strategies specifically aimed at optimizing escape routes for evacuees and first responders in wildfire scenarios. A survey paper on escape route planning analysis highlights that improved A* algorithms are safer and more reliable in wildfire scenarios~\cite{ZHU2024112355}. Notably, Liu et al.~\cite{10472809, liu2023active} utilize a fusion of satellite and UAV vision data to plan escape routes. They use weighted A* search that only incorporates the fire spread on the ground maps. Also, their approach does not consider a coordinated drone fleet deployment, nor does it account for resilience considerations. This omission is critical, as ensuring robustness and adaptability in UAV operations is essential for human-involved rescue missions, where safety is of paramount importance. In contrast, \textit{AeroResQ} integrates these crucial aspects, leveraging a distributed UAV network to provide real-time, resilient escape route planning.

\subsection{Resilience in UAV Fleet}
Using drone fleets with distributed UAV coordination for wildfire management has gained traction~\cite{bailon2022real}. However, it is crucial to incorporate resilience in multi-UAV operations~\cite{ordoukhanian2016resilient}. Seraj et al.~\cite{seraj2022multi} develop a cooperative planning approach that ensures continuous wildfire coverage and tracking, even in the face of dynamic fire behavior and UAV failures, thereby enhancing system resilience. Similarly, John et al.~\cite{10416753,10857313} introduce a decentralized sequential planner designed for early wildfire mitigation. They focus on resource efficiency and conflict awareness among heterogeneous UAV teams, which contributes to the robustness of wildfire response operations. \cite{HU2022107494} presents a fault-tolerant cooperation framework for UAV swarms in forest fire monitoring using graph-based navigation, decentralized task reassignment, and collision avoidance. In contrast, our work examines dynamic escape route planning and resilient coordination among UAVs for human evacuation, thus addressing real-time evacuation support.

\section{Discussion} \label{sec:discussion}

Deploying AeroResQ within government emergency-response systems poses several practical and organizational challenges. As highlighted in recent field studies on firefighters’ perceptions~\cite{10.1145/3613904.3642061}, effective integration must align with existing command hierarchies and communication workflows that are already well-defined and regulated. Moreover, limited technical familiarity and trust in autonomous systems among responders call for targeted training to foster confidence in drone-assisted operations. Since automated decisions may not always be perceived as fully reliable, AeroResQ may be initially used as a hybrid operational model that fuses drone-derived insights with data accessible and verifiable by firefighters on-site, ensuring transparency and human oversight in critical decision-making.

Further, scaling AeroResQ to a fleet of several hundred drones, let's say around $480$ drones ($400$ SDs and $80$ CDs) while maintaining the same SD:CD ratio preserves per CD workload, keeping SD-CD communication and computation bounded. The primary challenge lies in CD-CD coordination: as the number of CDs increases, inter-coordinator communication, state synchronization, and failure detection can quickly become bottlenecks. The current single-client centralized heartbeat mechanism must therefore evolve into a decentralized design, such as clustered coordination with local leaders, gossip-based heartbeats, or hierarchical federation across clusters. Since other AeroResQ components are already modular and decentralized, they are inherently suited to operate efficiently once the coordinator plane is extended to handle larger fleets.

\section{Conclusion and Future Work} \label{sec:conclusion}
In this paper, we introduced AeroResQ, a novel drone-based wildfire response system that enables collaborative path planning and real-time decision-making through a distributed on-device datastore and onboard deep learning models. By leveraging service drones and coordinator drones, AeroResQ efficiently detects stranded individuals, generates safe evacuation routes and monitors them until they reach their safe location. Our containerized emulation framework allowed for extensive evaluation under failure scenarios and fleet configurations, validating the scalability, efficiency, and resilience of our approach. Results on real wildfire datasets from Southern California (2025) demonstrated that AeroResQ achieves low-latency processing ($\leq500$ ms per request) making it a viable candidate for real-world deployments in large-scale wildfire emergencies.

As a part of future work, we plan to evaluate AeroResQ on dynamic wildfires and with complex cyber-infrastructure failure scenarios. Additionally, multi-UAV collaboration strategies will be further optimized to improve load balancing across the fleet. More sophisticated A* search algorithms can also be developed that can scale well with larger wildfires. In future, the platform can be extended with network resilience strategies that go beyond local WiFi or cellular networks, incorporating satellite communication links and ad hoc 5G/6G connectivity to ensure robust and reliable operation in remote or disaster-affected regions.


\section*{Acknowledgments}
This work was supported by the Department of
Energy Award \#DE-SC0024387, by the National Science Foundation Award \#2018074, and by the AI \& Robotics Technology Park (ARTPARK) at IISc. Suman Raj was supported by a Prime Minister's Research Fellowship, Ministry of Education, India.

\clearpage
\balance
\bibliographystyle{elsarticle-num} 
\bibliography{main}

\newpage
\nobalance
\appendix

\section{Guided Walk Simulation Strategy}\label{sec:guided-walk-strategy}

\begin{algorithm}[!t]
\caption{Guided Walk Simulation for Evacuee Movement}
\label{algo:guided-walk}
\begin{algorithmic}[1]
\Require Escape route waypoints $\lambda^{esc}_n$, current position $\lambda^f_n$, current time $t^f$, current index $ind_n$, update interval $\epsilon$, walking speed $v^f$
\Ensure $\lambda^f_n$ at $t^f + \epsilon$

\While{$status_n$ is $\texttt{ACTIVE}$}
    \State $d \gets v^f \times \epsilon$ \Comment{Compute expected displacement}
    \State $d_r \gets \text{HaversineDistance}(\lambda^f_n, \lambda^{esc}_n[ind_n])$ \Comment{Compute distance to next waypoint in the escape route} \label{L:3}

    \If{$d_r \leq d$} 
        \State $ind_n \gets ind_n + 1$ \Comment{Update waypoint index}
        \State Start back from step at line \ref{L:3}
    \Else
        \Comment{Waypoint is too far, move fractionally towards it}
        \State $d' \gets \text{HaversineDistance}(\lambda^f_n,\lambda^{esc}_n[ind_n])$
        \State $\gamma = (v^f \times \epsilon)/d'$ 
        \State $\lambda^f_n  \gets (1 - \gamma)\lambda^f_n + \gamma \lambda^{esc}_n[ind_n]$ \Comment{Compute and update next position}
        \State $t^f_n \gets t^f_n + \epsilon$ \Comment{Update timestamp}
    \EndIf
    
\EndWhile
\end{algorithmic}
\end{algorithm}

Since the evaluation of our proposed methodology is conducted through emulations, it becomes essential to accurately simulate the movement patterns of evacuees to reflect real-world behavior. To achieve this, we implement a \textit{guided walk simulation strategy}, which is formally described in Algorithm~\ref{algo:guided-walk}. The movement of a simulated evacuee must adhere to realistic constraints based on the natural walking speed of a person. According to empirical studies~\cite{Murtagh2020OutdoorWalkingSpeeds}, the average human walking speed is denoted as 1.5 meters per second. Given this information, we must determine the exact distance a virtual evacuee can cover within the interval between two consecutive location updates performed by the coordinator drone. We model the simulated firefighter (evacuee) as a proxy process.

Let $v^f$~m/s be the walking speed and $\epsilon$ represent the update interval duration, which is the fixed time interval at which the coordinator drone queries the evacuee's position. The process that serves as a proxy simulating the evacuee runs iteratively until the evacuee reaches the designated safe location. At each update instance, when the drone requests the evacuee's location, this proxy  retrieves the current GPS coordinates and timestamp of the firefighter's last known position. Let this last recorded location be $\lambda^f_n$ and the corresponding timestamp be $t^f_n$. 

The escape route assigned to an evacuee consists of predefined waypoints, which are essentially nodes in a spatial graph. These waypoints denote key path junctions rather than every minor step along the path. As a result, some waypoints may be placed far apart, covering long distances between two consecutive nodes. However, if an evacuee cannot traverse the full distance between two adjacent waypoints within the update interval $\epsilon$, the movement trajectory must be interpolated to track their intermediate positions accurately. To manage this interpolation, the proxy process maintains an index variable $ind_n$ that keeps track of the waypoint towards which the evacuee is currently headed. At any given time $t_0$, it calculates the expected position of the evacuee at $t_0 + \epsilon$ based on their walking speed $v^f$. This ensures that by the time the coordinator drone initiates the next location query at $t_0 + \epsilon$, it receives the precomputed position from the proxy, which accurately represents the firefighter's movement during the last interval.

\begin{figure}[!t]
    \centering
    \includegraphics[width=0.45\linewidth]{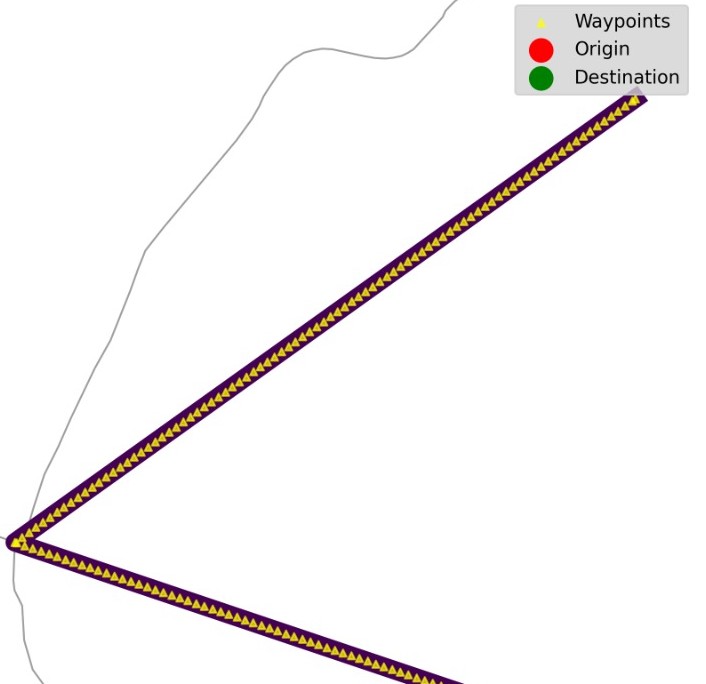}
    \caption{Guided walk visualisation}
    \label{fig:guided-walk-visualisation}
\end{figure}

An additional consideration arises when an evacuee traverses through a waypoint before the next scheduled update. If the firefighter moves a distance such that they cross a node waypoint from the escape route within a single interval $\epsilon$, the system needs to correctly update their progress. To address this, the proxy process implements a checkpoint validation mechanism: If the remaining geodesic distance calculated using Haversine formula between the evacuee’s current location and the next waypoint in the escape route (denoted as $\lambda^{esc}_n(ind_n)$), is less than the expected walking distance in that interval ($v^f \times \epsilon$), then we update the waypoint index to point to the next destination node in the sequence, i.e., $ind_n += 1$ and compute the evacuee's position from the current position to the updated value returned by $\lambda^{esc}_n(ind_n)$. 

The guided walk simulator operates independently on each coordinator drone, ensuring decentralized execution. Each coordinator drone runs a separate thread that has this proxy logic. It updates the last known location and timestamp exclusively for requests where the assigned processing coordinator drone ID matches the drone executing the thread. For instance, a thread running on \textit{cd-1} will handle simulations only for those evacuee requests whose $cID^p_n$ corresponds to \textit{cd-1}. For a specific escape route section, Fig.~\ref{fig:guided-walk-visualisation} visualizes the guided walk strategy, where the purple line represents the CD-generated route, and a series of very closely spaced yellow dots overlayed on the purple line indicate intermediate waypoints. This structured approach ensures that the movement of the simulated evacuee accurately reflects real-world firefighter's evacuation scenarios, enhancing both the reliability and effectiveness of the evaluation methodology.

\end{document}